\documentclass[twocolumn]{aastex63}
\usepackage[flushleft]{threeparttable}
\usepackage{lipsum}
\usepackage{placeins}
\usepackage{float}

\shorttitle{Abundances of Lensed Quiescent Galaxies at z $\sim$ 2}
\shortauthors{Jafariyazani et al.}

\begin{document}

\title{Chemical Abundances of Early Quiescent Galaxies: New Observations and Modelling Impacts}

\author[0000-0001-8019-6661]{Marziye Jafariyazani}
\affiliation{SETI Institute, Mountain View, CA, USA}
\affil{Carnegie Science Observatories, Pasadena, CA, USA }
\affil{Department of Physics and Astronomy, University of California, Riverside, CA, USA }

\author[0000-0001-7769-8660]{Andrew B. Newman}
\affil{Carnegie Science Observatories, Pasadena, CA, USA }

\author[0000-0001-5846-4404]{Bahram Mobasher}
\affil{Department of Physics and Astronomy, University of California, Riverside, CA, USA }

\author[0000-0002-5615-6018]{Sirio Belli}
\affil{Dipartimento di Fisica e Astronomia, Università di Bologna, Bologna, Italy}

\author[0000-0001-7782-7071]{Richard S. Ellis}
\affil{Department of Physics and Astronomy, University College London, Gower Street, London WC1E 6BT, UK}

\author[0000-0002-9382-9832]{Andreas L. Faisst}
\affiliation{Caltech/IPAC, 1200 E. California Blvd. Pasadena, CA 91125, USA}

\begin{abstract}

Recent stellar chemical abundance measurements of a handful of $z\sim2$ quiescent galaxies have suggested these galaxies exhibit a remarkably strong $\alpha$-enhancement compared to their local and intermediate redshift counterparts. This apparent chemical evolution following quenching suggests that even the innermost regions of massive early-type galaxies may have experienced substantial mixing of stars in mergers, challenging a purely inside-out growth model. However, larger samples are needed to determine whether a high $\alpha$-enhancement ([Mg/Fe] $\approx 0.5$) is common in $z \sim 2$ quiescent galaxies, and a comparative analysis is needed to determine whether it is consistently inferred using different stellar population synthesis models. We report age and stellar chemical abundance measurements for a sample of four gravitationally lensed quiescent galaxies at $z\sim2.1-2.65$ based on Magellan/FIRE spectroscopy. For three of these galaxies we constrain the $\alpha$-enhancement, and in two cases we measure high values comparable to earlier results when the spectra are analyzed consistently. We also find that the choice of modeling approach can exert a significant effect on the measured abundances. This model dependence can be partly, but not entirely, explained by the complex abundance patterns of $\alpha$ elements in galaxies, which has been observed at lower redshifts and in one $z \sim 2$ quiescent galaxy. Our investigation highlights the importance of independently varying abundance of $\alpha$ elements when fitting the spectra of such galaxies. Observations with JWST will soon deliver precise and spatially resolved abundances of these and other quiescent galaxies at cosmic noon, opening a new window into their evolution.

\end{abstract}


\section{Introduction} \label{sec:intro3}

Among the key outstanding questions in the study of galaxy formation and evolution are the processes resulting in the quenching of star formation. These processes can be studied from different perspectives using different probes and galaxy populations. To address this question, ultimately one needs to reconstruct the formation history of quiescent galaxies. One way to explore this is to investigate stellar chemical abundances of quiescent galaxies using diagnostics such as the $\alpha$-enhancement [$\alpha$/Fe], specifying the $\alpha$-element-to-iron ratio. [$\alpha$/Fe] is believed to be sensitive to the star formation timescale because core-collapse supernovae (SN) produce mostly $\alpha$-elements (e.g., oxygen, magnesium, silicon, etc., \citealt{Woosley1995}) quickly, whereas SN Type Ia mostly release iron peak elements over a range of timescales extending to many Gyr. This diagnostic has been extensively used to deduce the star formation history (SFH) of local early-type galaxies (ETGs; e.g., \citealt{Thomas2003,Greene2019}). 

However, galaxies undergo minor and major mergers throughout their history. These mergers mix in stars that were born in various galaxies with different SFHs, which complicates the interpretation of local archaeological studies. Therefore we must also measure the chemical abundances and SFHs of quiescent galaxies at high redshifts around the time they experienced quenching. These systems have been challenging to observe spectroscopically, especially from the ground. To date, there are two quiescent galaxies at $\textit{z} \sim$ 2 for which $\alpha$-enhancement and iron abundance have been measured with $\sim$ 0.1 dex precision: COSMOS-11494 (\citealt{Kriek2016}) and MRG-M0138 (\citealt{jafari2020}). COSMOS-11494 and MRG-M0138 were both found to have unusual abundances compared to local ETGs. Specifically, their high [Mg/Fe] ($0.59\pm0.11$ for COSMOS-11494 and $0.51\pm0.05$ for MRG-M0138) suggested a very short formation timescale based on simple chemical evolution models. 

In a recent study by \citet{Beverage2023}, [Mg/Fe] and [Fe/H] were robustly measured for 10 massive quiescent galaxies at z $\sim$ 1.4 and four at z $\sim$ 2.1 using spectra from Keck/MOSFIRE. They found subsolar [Fe/H] with an average value of approximately $-0.2$ for z $\sim$ 1.4 galaxies and $-0.3$ for their z $\sim$ 2.1 sample. They found an average [Mg/Fe] $\approx 0.3$ at $z \sim 1.4$, similar to local ETGs, increasing to [Mg/Fe] $\approx 0.5$ at z $\sim$ 2.1. Like the initial measurements, these findings suggest extremely short star-formation timescales for z $\sim$ 2.1 quiescent galaxies.

The chemical dissimilarities that have been observed between $z\sim2$ massive quiescent galaxies and local ETGs suggests that a significant fraction of the stars in the very centers of today's ETGs were formed \emph{ex situ} and delivered in mergers long after quenching. It appears that even the cores of massive ellipticals did not evolve passively, but instead experienced surprisingly complex evolution. Therefore, it is important to understand whether the stellar chemical compositions of the few examples observed to date are typical of the $z\sim2$ quiescent population. This requires more observations, and also an investigation of the degree to which the high-$z$ measurements are dependent on the underlying stellar population models.

One of the aforementioned galaxies, MRG-M0138 \citep{jafari2020}, was among a unique sample of five gravitationally lensed, massive, quiescent systems at $ 1.95 < \textit{z} < 2.64 $ first presented in \cite{Newman2018a}. In this paper, we use the same methods employed by \citet{jafari2020} to investigate the stellar populations of the additional four lensed quiescent galaxies: MRG-M2129, MRG-P0918, MRG-S1522 and MRG-M0150. Using deep ground-based spectroscopic observations of these bright lensed galaxies, we measure the age, [Fe/H], and in most cases [$\alpha/$Fe] to offer new and complimentary constraints on the formation of these early quiescent systems. Furthermore, we model our spectra with multiple stellar population synthesis models to investigate the potential effects of modeling choices on stellar chemical abundance measurements.

This paper is organized as follows. In Section \ref{sec:data3} we briefly describe our data. In Section \ref{sec:spec-fit} we explain the procedure and assumptions to measure chemical abundances from our spectra using two sets of stellar population synthesis models, and we present our measurements. In section \ref{sec:model-comparison} we compare the results of using different models and address the reasons for some of the discrepancies. We discuss the interpretation of our results for galaxy evolution models in section \ref{sec:chem_abundance} and summarize our findings in section \ref{sec:summery}. 

\begin{figure*}
\includegraphics[width=\textwidth]{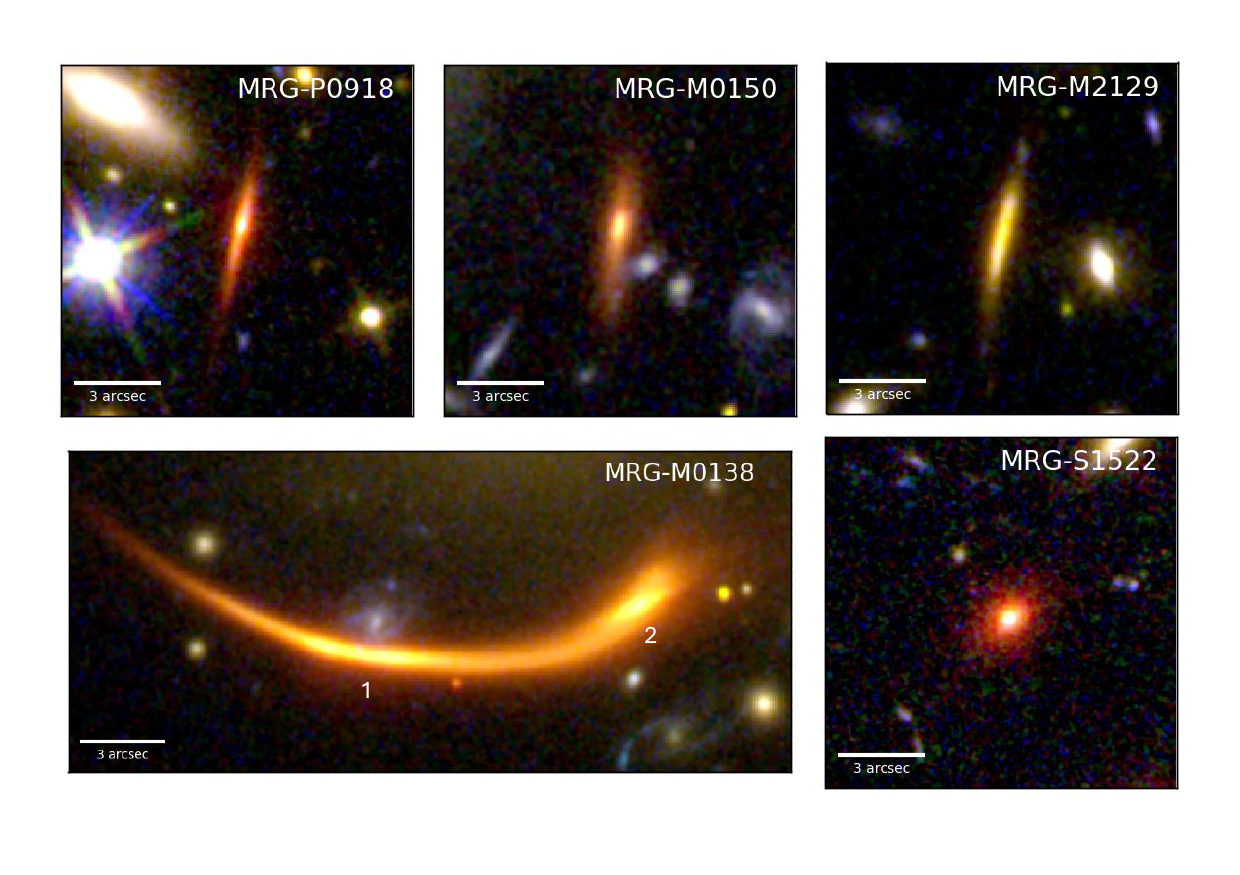}
\caption{HST images of our sample of lensed quiescent galaxies. See \protect\citet{Newman2018a} for wider-field images showing the lensing clusters and multiple images of the targets, as well as details of the imaging data. Images are displayed with a logarithmic stretch using one of the F555W, F606W, or F814W filters for the blue channel; F105W or F125W for the green; and F160W for the red. For MRG-M0138, two images of the galaxy are labeled as 1 and 2. The spectrum from image 1 is from Keck/MOSFIRE, while the spectrum from image 2 is from Magellan/FIRE.}
\label{fig:image}
\end{figure*}

\begin{figure*}
\includegraphics[width=0.95\textwidth]{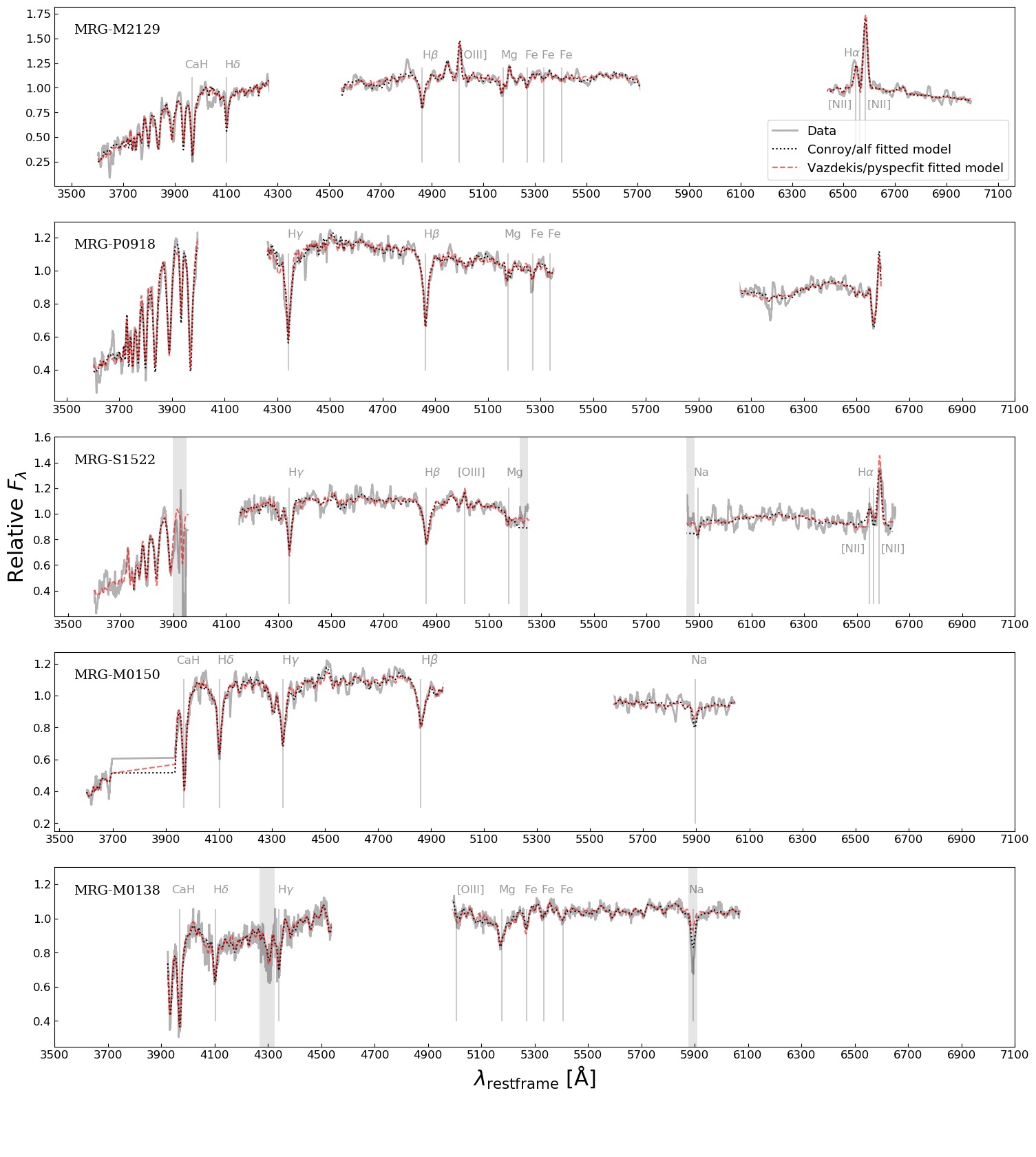}
\caption{Each panel presents the integrated spectra of individual galaxies in grey which are all taken from Magellan/FIRE, except for MRG-M0138 in the last panel which is taken by Keck/MOSFIRE. The best-fit model employing Conroy models fitted with \emph{alf} is shown in dotted black line, and the best fit Vazdekis models fitted with \emph{pyspecfit} are shown in dashed red line for each galaxy. Some prominent spectral lines for each of the spectra are also shown. Grey shaded regions are masked in the spectral fitting process. For visualization purposes, all FIRE spectra are smoothed by 600 km s$^{-1}$ using inverse variance weighting, and the relative fluxes are presented in the y-axes and do not begin from 0.}
\label{fig:spectra}
\end{figure*}

\section{Data} \label{sec:data3}

In \citet{jafari2020}, we presented a comprehensive study of MRG-M0138 using Keck/MOSFIRE (\citealt{McLean2010_MOSFIRE,MOSFIRE}) and Magellan/FIRE (\citealt{FIRE}) spectra. In this paper we investigate the stellar chemical abundances of MRG-2129, MRG-M0150, MRG-P0918 and MRG-S1522, the other lensed quiescent galaxies introduced by \citet{Newman2018a},\footnote{MRG-M2129 was discovered by \citet{geier2013}.} who also provided a full description of each galaxy and the relevant spectroscopic observations. Figure \ref{fig:image} also presents Hubble Space Telescope (HST) imaging of this sample. 

Our analysis is based on Magellan/FIRE spectra. For three of the four galaxies (all but MRG-S1522), these spectra are spatially resolved. \citet{Newman2018/2} used the resolved spectra to investigate the rotation of the systems and here, as in \citet{jafari2020}, we remove the resolved rotation measured by \citet{Newman2018/2} when constructing the integrated light spectrum of each galaxy. Specifically, we shift each row of the two-dimensional spectrum to the galaxy systemic redshift, before extracting an integrated light spectrum using optimal weighting to maximize the signal-to-noise ratio. Since the rotation speeds are much higher than the FIRE spectral resolution, this procedure has the advantage of significantly reducing sky subtraction residuals, which often dominate near-infrared ground-based spectra.

The median signal-to-noise ratio (SNR) in the $H$ band ranges from 15 to 25 per pixel in our 25 km s${}^{-1}$ bins. In 250 km s${}^{-1}$ bins that are better matched to the stellar velocity dispersion, this SNR ranges from 49 to 79. These are quite high SNRs for spectra of $z \gtrsim 2$ galaxies observed at comparable spectral resolution, which is a consequence of the lensing magnification. 

\section{Spectral Fitting} \label{sec:spec-fit}

To measure stellar chemical abundances from these spectra, we first use the stellar population synthesis (SPS) models from \citet{Conroy2012a}, as updated by \citet{Conroy2018}, along with the \emph{alf} fitting code \citep{conroy_alf2023}. This is consistent with our previous work on MRG-M0138 \citep{jafari2020}. We further model these four spectra using the \citet{vazdekis2010,vazdekis2015} SPS models to compare the effects of utilizing different modeling approaches. We choose these models because they enable full-spectral fitting with variable $\alpha$-enhancement, and as discussed by \citet{Kriek2016}, full-spectrum fitting is far preferable to traditional index analyses for ground-based near-infrared spectra. Here we fit the \citet{vazdekis2010,vazdekis2015} models to the spectra using the \emph{pyspecfit} code \citep{newman2014} as described below.

\subsection{Conroy SPS models and alf} \label{sec:alf}
In our first modeling approach we utilize \emph{alf} to model the spectra with simple stellar populations from the \citet{Conroy2018} suite, which in turn are based on a combination of empirical and synthetic stellar libraries (\citealt{Sanchez-Blazquez2006,choi2016,Villaume2017}). The code can be run in three different modes depending on the quality of data and the desired complexity of the model. Although our spectra do not have sufficient SNR to constrain all the parameters included in the full fitting mode, we choose this mode for our analysis to be able to take into account the emission lines and to rescale errors by a jitter term, features that are only available in this mode. 

Our basic input assumptions are a Kroupa initial mass function (IMF, \citealt{Kroupa}) as well as the following limits for the abundance of individual elements: $-0.3 < [X/H] < +1$. Our final results are based on a single-age  population model; however, we also performed the fit assuming a two-component star formation history which includes a younger and a relatively older component within the galaxy, and the resulting abundances remained consistent with the single-component model. Our spectra and their corresponding best-fit models are shown on Figure \ref{fig:spectra} in grey and black dotted line, respectively, and the measured age, [Mg/Fe] and [Fe/H] are listed in Table \ref{table:alf}. We did not include the [Mg/Fe] measurement for MRG-M0150 in our analysis as its spectrum does not cover the key Mg b lines (at 5167, 5172 and 5183 \AA). Also, we caution that the lower limit of age in these stellar population models is 1 Gyr, and for ages between 0.5 to 1 Gyr \emph{alf} will extrapolate off the model grid. In our sample this affects only the youngest galaxy, MRG-P0918,  with an estimated stellar population age of $0.65\pm0.07$ Gyr; therefore, the results for this galaxy should be interpreted with caution.

\begin{figure*}
\centering
\includegraphics[width=0.7\textwidth]{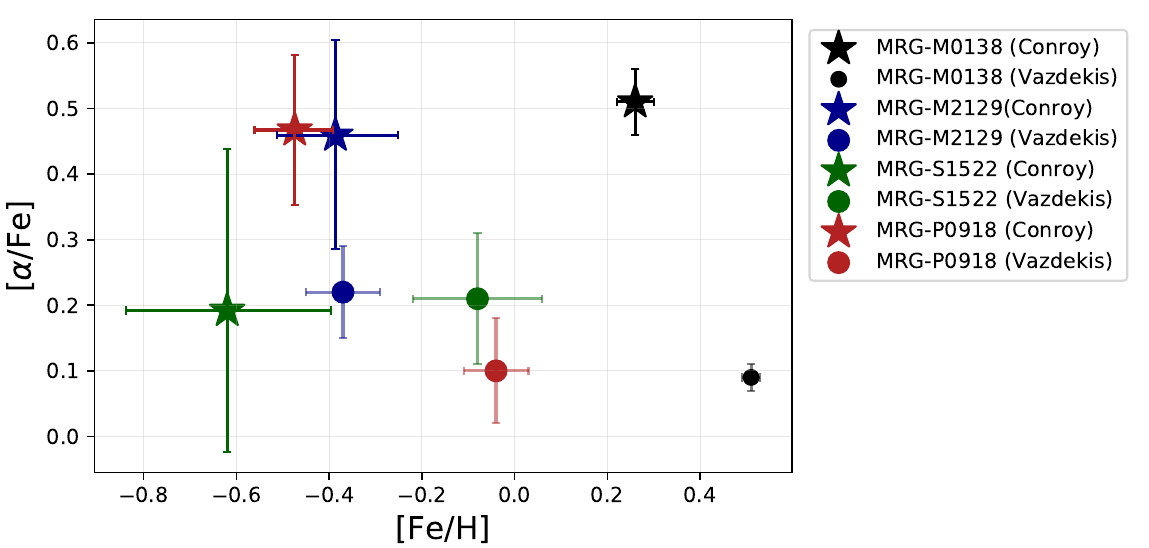}
\caption{Comparison of abundances derived from the Conroy and Vazdekis stellar population synthesis models using \emph{alf} and \emph{pyspecfit}, respectively. Each color represents a single galaxy, with star markers indicating Conroy/\emph{alf} results and circle markers indicating Vazdekis/\emph{pyspecfit} results.}
\label{fig:conroy_vazdekis}
\end{figure*}

\startlongtable
\begin{deluxetable*}{lccc}
\tablecaption{Stellar Population Parameters \label{table:alf}}
\tablehead{
\colhead{Galaxy (Model)} & \colhead{Age (Gyr)} & \colhead{[Mg/Fe]} & \colhead{[Fe/H]}
}
\startdata
MRG-M2129 (Conroy) & $1.05\pm0.2$ & $0.58\pm0.21$ & $-0.29\pm0.21$ \\
MRG-M2129 (Vazdekis) & $1.15\pm0.09$ & $0.22\pm0.07$ & $-0.37\pm0.08$ \\
\hline
MRG-P0918 (Conroy) & $0.65\pm0.07$ & $0.40\pm0.15$ & $-0.49\pm0.12$ \\
MRG-P0918 (Vazdekis) & $0.5\pm0.03$ & $0.1\pm0.08$ & $-0.04\pm0.07$ \\
\hline
MRG-S1522 (Conroy) & $1.08\pm0.17$ & $0.19\pm0.23$ & $-0.62\pm0.22$ \\
MRG-S1522 (Vazdekis) & $0.64\pm0.06$ & $0.21\pm0.1$ & $-0.08\pm0.14$ \\
\hline
MRG-M0150 (Conroy) & $1.28\pm0.34$ & N/A* & $-0.56\pm0.33$ \\
MRG-M0150 (Vazdekis) & $0.87\pm0.06$ & N/A* & $-0.22\pm0.17$ \\
\hline
MRG-M0138 (Conroy) & $1.37\pm0.11$ & $0.51\pm0.05$ & $0.26\pm0.04$ \\
MRG-M0138 (Vazdekis) & $0.9\pm0.01$ & $0.09\pm0.02$ & $0.51\pm0.02$ \\
\enddata
\tablecomments{*Mg b lines are not covered in the spectra of this galaxy for a reliable [Mg/Fe] measurement.}
\end{deluxetable*}

\begin{figure*}
\includegraphics[width=\linewidth]{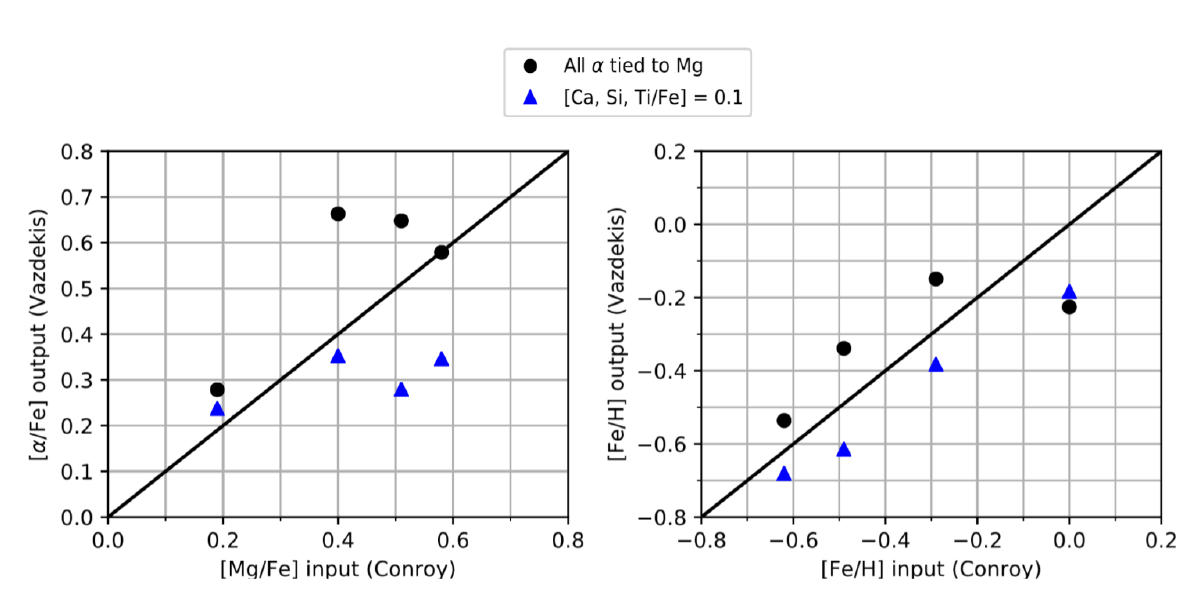}
\caption{Comparison of abundances inferred by fitting the Vazdekis models to spectra generated with the Conroy models. For each observed galaxy (except MRG-M0150), we generate Conroy models with matched age and [Fe/H] and fit Vazdekis models to the spectra. Black points show models with the abundance of all $\alpha$ elements (see list in text) fixed to the measured [Mg/Fe], while for the blue points Ca, Si, and Ti are instead set to [X/Fe] = 0.1. The [$\alpha$/Fe] inferred with the Vazdekis fits depends strongly on the input $\alpha$ abundance pattern.
\label{fig:modelcomp}}
\end{figure*}

\subsection{Vazdekis SPS models and pyspecfit} \label{sec:pyspecfit}

We additionally model each of these four spectra, as well as the Keck/MOSFIRE spectra of MRG-M0138 \citep{jafari2020}, using a grid of SPS models from \citet{vazdekis2015}, which built upon the \citet{vazdekis2010} library by adding an $\alpha$-enhanced grid with $[\alpha/{\rm Fe}] = +0.4$. As for the Conroy models, we fit simple stellar populations (SSPs), but the Vazdekis SSPs are computed to much younger ages. Unlike the Conroy models, the Vazdekis models tie together the abundances of a defined set of $\alpha$ elements (O, Ne, Mg, Si, S, Ca, and Ti) rather than varying them independently, which is important to bear in mind when comparing the results. We use the models with a \citet{chabrier} IMF.

We use the Bayesian code \emph{pyspecfit} to infer the stellar population parameters from the data using the Vazdekis models. The free parameters are the redshift, stellar velocity dispersion, SSP age, [Z/H], [$\alpha$/Fe], and a set of parameters describing the emission line velocities, widths, and intensities as described in \citet{newman2014} (the exception is MRG-M0138, for which we see no evidence of line emission and so do not model it). We linearly interpolate the spectra in log age, [Z/H], and [$\alpha$/Fe] in order to sample models from a continuous grid. Note that the maximum [Z/H] is +0.4 and the only [$\alpha$/Fe] values are 0 and +0.4; we allow a linear extrapolation beyond these values in the fitting, but ultimately find that the posteriors are not very close to these limits.

From these model parameters we also compute $[{\rm Fe/H}] = [{\rm Z/H}] - 0.75 [\alpha/{\rm Fe}]$ as specified by \citet{vazdekis2015}.
Similar to our Conroy/\emph{alf} analysis, we divide the spectra into wavelength chunks corresponding to the atmospheric transparency windows, and within each chunk, we determine the multiplicative polynomial that best matches the spectral shape of the proposed model to that of the data. We modify the spectra with this polynomial before computing the likelihood (see the appendix of \citealt{newman2014}). The polynomial order is approximately the wavelength range divided by 200~\AA. Thus, our abundance measurements are insensitive to the continuum shape, which is affected by dust attenuation and flux calibration uncertainties, on these and larger scales. Finally, we perform small multiplicative rescalings of the error spectrum so that the best-fitting model has $\chi^2 / {\rm dof} \approx 1$.

The best-fit Vazdekis/\emph{pyspecfit} models to our spectra are also shown on Fig. \ref{fig:spectra} in red dashed line and the measured age, [Mg/Fe] and [Fe/H] are also presented in Table \ref{table:alf}.

\section{Understanding model differences} \label{sec:model-comparison}

In Figure~\ref{fig:conroy_vazdekis}, we compare abundances of each galaxy derived using the Conroy and Vazdekis SPS models. The Vazdekis-based [Fe/H] are systematically higher by 0.3 dex on average, while the [$\alpha$/Fe] are systematically lower than Conroy-based [Mg/Fe] by 0.3 dex. Also, the Vazdekis-based ages are systematically younger by 30\%.

Understanding these differences is obviously critical, since their magnitude is large and could lead to very different physical interpretations. One key difference between the Conroy and Vazdekis models is that the abundances of the $\alpha$ elements O, Ne, Mg, Si, S, Ca, and Ti are tied together in the Vazdekis models, whereas in our Conroy/\emph{alf} fitting procedure the abundances of elements are varied independently, and we report [Mg/Fe] specifically. In ETGs at $z \sim 0$ and 0.7, the $\alpha$ elements heavier than Mg are found to be less enhanced \citep{Conroy2014,Choi2014,Beverage2023a}. The underabundance of calcium in particular was long regarded as a puzzle \citep{saglia2002,Thomas_Ca,Parikh2019}. In MRG-M0138 at $z\sim2$, \citet{jafari2020} found that both Si and Ca were much less enhanced ([Si, Ca/Fe] $\approx 0.1$) than Mg ([Mg/Fe] $\approx 0.5$).

We perform a simple model comparison to test whether the $\alpha$ abundance pattern could underlie the apparent differences in our results based on the Conroy and Vazdekis models. We first generate Conroy spectra that have parameters similar to the observed sample: matched age, [Fe/H], and [$\alpha$/Fe] set to the observed [Mg/Fe] for all the $\alpha$ elements listed above.\footnote{For the MRG-M0138-like spectrum, we instead set [Fe/H] = 0 to avoid exceeding the maximum metallicity in the Vazdekis grid.} We then fit the Conroy spectra with the Vazdekis models and compare the recovered parameters to the inputs. This is a direct comparison of consistency when the models use the same abundance patterns.

We find that the input [$\alpha$/Fe] and [Fe/H] are recovered reasonably well (black points, Figure~\ref{fig:modelcomp}), at the 0.1 dex level in 3 of 4 cases. (The largest outlier is the [$\alpha$/Fe] of MRG-P0918, the youngest galaxy, where we expect larger model uncertainty.) We note that this test requires extrapolating the Vazdekis models above the grid limit [Mg/Fe] $ = 0.4$. Nonetheless, the key point is that we do not infer systematically lower [$\alpha$/Fe] with the Vazdekis models, as we do in the observed galaxies, once the $\alpha$-abundance patterns are matched. In fact, the recovered [$\alpha$/Fe] are slightly higher than input. We then generated a second set of spectra in which we modified the abundances of the heavier $\alpha$ elements such that [Ca, Si, Ti/Fe] = 0.1, which approximates the results from MRG-M0138 and local ETGs. The blue triangles in Figure~\ref{fig:modelcomp} show that the inferred [$\alpha$/Fe] drops substantially, in most cases by 0.2-0.3 dex, roughly consistent with the difference between the Conroy- and Vazdekis-based abundance ratios in the observations.
 
This suggests that the detailed $\alpha$ abundance pattern, accounting for the independent variation of the $\alpha$ elements, may well be the origin of the apparent discrepancies in [Mg/Fe] (Conroy) versus [$\alpha$/Fe] (Vazdekis) in Table~\ref{table:alf} and Figure~\ref{fig:conroy_vazdekis}. However, this test does not reproduce the comparably large [Fe/H] discrepancies, which suggests there could be deeper model differences in the metallicity scale at these ages. 

\section{The Evolution of Stellar Chemical Abundances in Quiescent Galaxies} \label{sec:chem_abundance}

\begin{figure*}
\includegraphics[width=\textwidth]{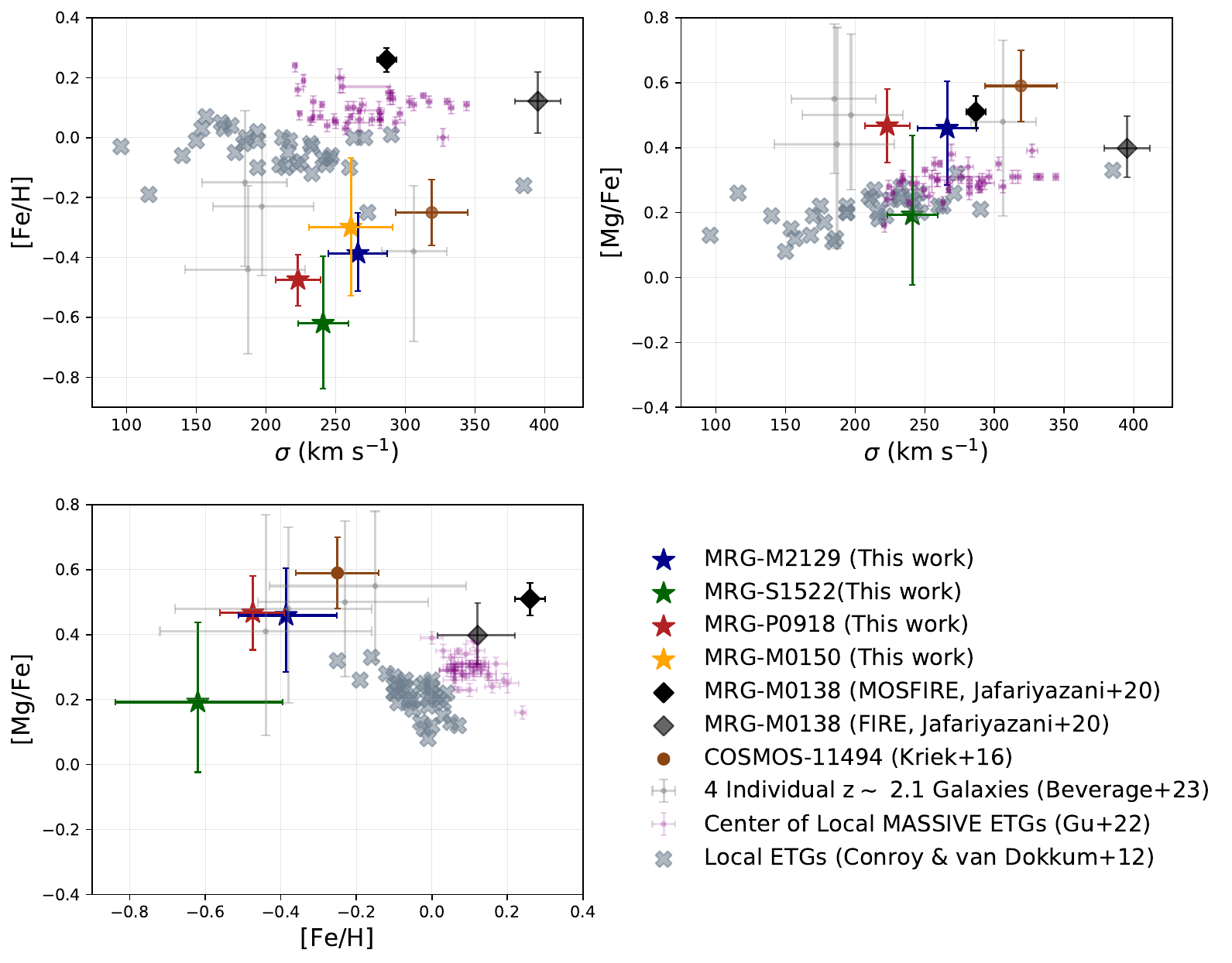}
\caption{Abundance patterns of $z\sim2$-2.6 quiescent galaxies compared to their local counterparts. Top left: [Fe/H] versus velocity dispersion ($\sigma$), top right: [Mg/Fe] versus velocity dispersion, lower left: [Mg/Fe] versus [Fe/H]. The sample of $z \sim2$-2.6  quiescent galaxies includes the four galaxies analyzed in this paper (MRG-M2129, MRG-S1522, MRG-M0918, and MRG-M0150. For MRG-M0150, only [Fe/H] is reported because the Mg b lines are not covered in the spectra of this galaxy. We also included six galaxies at $z\sim2$ from the literature: MRG-M0138 (black and gray diamonds based on the Keck/MOSFIRE and Magellan/FIRE spectra, respectively; \citealt{jafari2020}), COSMOS-11494 (brown circle, \citealt{Kriek2016}) and four galaxies from \citep{Beverage2023} in the grey circle datapoints. The local ETG sample includes ETGs from \citealt{Conroy2012a} (grey filled X) and the centers of MASSIVE ETGs in light purple circles from \citep{gu2022}. z $\sim2$-2.6 galaxies, except MRG-M0138, exhibit lower Fe/H compared to typical local ETGs, and they are typically more Mg-enhanced.}
\label{fig:mg/fe}
\end{figure*}

Our tests in Section~\ref{sec:model-comparison} highlight the importance of independently varying the abundances of $\alpha$ elements, such as O, Mg, Si, S, Ca, and Ti, when performing full spectral fitting. Therefore we proceed with our analysis using the results obtained from fitting the spectra with the Conroy SPS models. This also allows a direct comparison with other galaxies at $z\sim2$ and $z\sim0$ that have been analyzed in the same way. 

Figure \ref{fig:mg/fe} presents our four new measurements alongside the results for the two analogous systems with the most precise measurements ($\sim$ 0.1 dex precision), COSMOS-11494 \citep{Kriek2016} and MRG-M0138 \citep{jafari2020}, and for four unlensed massive quiescent galaxies at $z\sim2.1$ from \citet{Beverage2023}. We also overlay two different samples of local early-type galaxies from \citet{Conroy2012a} and \citet{gu2022}. The sample from \citet{gu2022} included stellar population measurements solely for the central region of ETGs, within an effective circular radius of R$_{e}$/8, enabling us a comparison specifically with the inner regions of local galaxies. Such a comparison is appropriate because the $z > 2$ galaxies are thought to evolve into the cores of massive ellipticals; we expect that moderate variations to the aperture would have a mild effect on [Fe/H] and none on [Mg/Fe] \citep{gu2022}. The top panels of Figure \ref{fig:mg/fe} demonstrate [Fe/H] and [Mg/Fe] values as a function of velocity dispersion ($\sigma$), and the bottom panel demonstrates their distribution in the [Mg/Fe] versus [Fe/H] parameter space.

Figure \ref{fig:mg/fe} shows that all of the four newly analyzed galaxies have markedly low [Fe/H] compared to local ETGs, consistent with the low [Fe/H] of COSMOS-11494 and the four galaxies from \citet{Beverage2023}, but significantly distinct from MRG-M0138, which has a remarkably high [Fe/H]. In the [Mg/Fe] versus velocity dispersion parameter space, we have three new measurements since, as discussed earlier, the MRG-M0150 spectrum does not cover the Mg b lines, limiting our ability to accurately constrain its [Mg/Fe]. Of these three new measurements, MRG-M2129 and MRG-P0918 are similar to the previous $z \sim 2$ measurements, all of which are Mg-enhanced compared to typical local ETGs with ${\rm [Mg/Fe]} \approx 0.4-0.6$. MRG-S1522 nominally has [Mg/Fe] more comparable to local ETGs and could have evolved into a typical local ETG with no merger-driven evolution, but its large uncertainty also can make it consistent with Mg-enhancement seen in the other $z\sim2$ systems. 
Our results suggest that enhanced [Mg/Fe] ratios, compared to local ETGs, are likely a widespread phenomenon among early quenched galaxies. If these galaxies evolved passively, their abundance pattern would be frozen in, and one would expect abundances at $z \sim 2$ to match those of their $z=0$ descendants. If the descendants of our sample have abundance ratios typical of local massive ETGs, then our Conroy-based measurements would instead imply a decline in Mg/Fe of roughly $2\times$ ($0.3\,{\rm dex}$) on average.

As discussed by \citet{Kriek2016}, minor mergers can be expected to reduce a galaxy's [Mg/Fe] ratio after quenching, and they present a simple model that could explain the observed evolution. This is plausible, although it is somewhat surprising that the mergers so strongly affect the central abundances, whereas the conventional interpretation of the surface brightness profile evolution is that minor mergers mainly deposit stars at large radii \citep{vanDokkum2010,hilz2013}. Furthermore, no evolution in Mg/Fe has been observed after $z \sim 0.7$ \citep{Choi2014,Bevacqua2023}, which significantly shortens the time available for mergers to dilute Mg/Fe.

Progenitor bias effects may also contribute but are not likely the whole story. In this picture, the descendants of galaxies that quenched the earliest would be found in the tails of the [Fe/H] and [Mg/Fe] distributions seen at lower redshifts. The problem, as discussed by \citet{Beverage2023}, is that we do not observe any analogs in the cores of local massive ETGs in the MASSIVE survey (Fig.~\ref{fig:mg/fe}). Larger samples of homogeneously measured stellar abundances across redshift could enable a number density analysis to better constrain the role of progenitor bias in chemical evolution.

There are also other interesting challenges posed by the high [Mg/Fe] ratios. Since they approach or even exceed the plateau value of ${\rm [Mg/Fe]} \approx 0.4$ seen in Galactic low-metallicity stars \citep{McWilliam1995,McWilliam2016}, they imply enrichment almost entirely by core collapse supernova  with very little contribution from SN Ia. In a conventional interpretation based on simple chemical evolution models, this leads to the inference of a short star formation timescale: in the case of COSMOS-11494, \citet{Kriek2016} find $\lesssim 500$~Myr. This in turn implies a past average star formation rate of $\gtrsim 500$ $M_{\odot}$ yr${}^{-1}$ during the star-formation phase of the galaxies in our sample. Since the high (Conroy-based) [Mg/Fe] ratios appear to be common in massive quiescent galaxies at $z\sim2$, this would link most of these galaxies to an early starburst phase; $z > 3$ sub-millimeter galaxies would be good candidate progenitors (e.g., \citealt{Toft2014}). Although a chemical link to ancient starbursts is enticing, there are also signs that the conventional timescale interpretation may be too simple. In particular, simple chemical evolution models cannot produce both the high [Mg/Fe] and super-solar [Fe/H] seen in MRG-M0138, which might require additional ingredients such as a flatter high-mass IMF \citep{jafari2020}. 

Finally, we have taken care to compare high- and low-redshift samples that were analyzed consistently with the Conroy SPS models and the \emph{alf} code, which in principle reduces the systematic uncertainties in such comparisons. Nonetheless, the stellar ages are of course very different in the $z\sim2$ and local samples. It remains possible that systematic uncertainties in SPS models may contribute to the apparent chemical evolution.

\section{Discussion and Summary} \label{sec:summery}

We report stellar chemical abundances of four gravitationally lensed quiescent galaxies at $z=2.1-2.65$, expanding the sample of the six comparable systems analyzed in the literature. We measure [Fe/H] and, in three cases, [Mg/Fe] as a proxy of $\alpha$ enhancement for this sample. These constraints are enabled by the high signal-to-noise ratio of our spectra (Section~\ref{sec:data3}) due to the lensing magnification. A key result is that at least 2 of the 4 galaxies (MRG-M2129 \& MRG-P0918) show highly elevated [Mg/Fe] ratios, comparable to those previously reported by \citet{Kriek2016}, \citet{jafari2020}, and \citet{Beverage2023}, when the data are analyzed similarly using the Conroy SPS models. The third galaxy (MRG-S1522) could also be consistent with an elevated [Mg/Fe], given its larger error bar. The spectrum of the fourth galaxy (MRG-M0150) did not permit the measurement of [Mg/Fe].

Our new measurements reinforce the pattern that high-$\textit{z}$ quiescent galaxies typically exhibit high [Mg/Fe] ratios and low [Fe/H] abundances compared to the cores of local ETGs with similar velocity dispersions. The very high [Mg/Fe] values could be indicative of a very short formation timescale ($\lesssim 200$~Myr, e.g., \citealt{Beverage2023}) according to conventional chemical evolution models in which core-collapse SN produced most of $\alpha$-elements, while longer lived SN Type Ia were not yet able to increase the iron abundance. At the same time, the lack of chemical analogs of the $z\sim2$ galaxies among the cores of today's ellipticals suggests that massive galaxies continued to evolve chemically after quenching, perhaps through mergers that mixed stars well into the galaxy cores.

We further analyzed our spectra using two different SPS models that allow for variable $\alpha$-enhancement. Our analysis shows that Vazdekis SPS models give substantially lower (0.3 dex) [$\alpha$/Fe] and higher [Fe/H] for these galaxies compared to the Conroy SPS models. At face value, this was an alarming discrepancy, since it is large enough to change the qualitative conclusions, e.g., the galaxies would instead be consistent with passive chemical evolution. However, as discussed in Section~\ref{sec:model-comparison}, the $\alpha$-enhancement seems to vary rather widely among $\alpha$ elements in studies of $z \sim 0$-0.7 ETGs as well as MRG-M0138. We showed that a plausible $\alpha$-abundance pattern could lead to an inferred [$\alpha$/Fe] based on the Vazdekis models, which boost all $\alpha$ elements equally, that is about 0.2-0.3 dex smaller than [Mg/Fe]. The tests in Section~\ref{sec:model-comparison} suggest that the apparently large differences in [$\alpha$/Fe] (Vazdekis) versus [Mg/Fe] (Conroy) may not reflect a genuine discrepancy, but rather abundance variations among the $\alpha$ elements. The [Fe/H] differences, on the other hand, are large (0.3 dex) and are not readily explained by our tests. These might reflect genuine model systematic uncertainties which require further investigation.

It should also be noted that there is some evidence from emission line ratios and widths of shocked gas and/or AGN in our sample \citep{Newman2018a}. Recent JWST unresolved spectra of quiescent galaxies at z $\sim$ 2--3 have found that such emission line patterns are common \citep{bugiani2024}, as are neutral gas outflows \citep{belli2024,Davies2024}; these probably play an important role in the chemical evolution of this population. Upcoming JWST IFU data will allow us to spatially map gas emission and absorption, enabling a more detailed investigation.

Our findings emphasize that a comprehensive analysis incorporating chemical evolution models, detailed star formation history reconstructions, and consideration of structural evolution is needed to explain the evolution of quiescent galaxies consistently. Such models could inform several parameters affecting the chemical evolution, including core-collapse supernova yields, SN Type Ia delay time distributions, gas flows, and the shape of the IMF.

~\emph{JWST} observations of high-$z$ quiescent galaxies will soon provide multi-element stellar abundance patterns able to constrain such models, as already presented by \citet{jafari2020} for a single lensed galaxy. MRG-M0138 has been recently observed by \emph{JWST} as part of the Cycle 1 proposal GO-2345, and the four galaxies presented in this paper will be observed as part of the Cycle 3 proposal GO-4903. These observations will offer a remarkably detailed and spatially resolved perspective to understand the variations of the abundances discussed in this paper within the galaxies and their implications on our physical models. Also they will lay the groundwork for future studies on statistically large samples of high-$z$ quiescent galaxies, particularly those magnified by gravitational lenses. This will enable far more detailed studies and will be achievable when Euclid \citep{Euclid2011,euclid2013} and the Nancy Grace Roman Space Telescope \citep{wfirst2015,roman2019} discover thousands of strongly lensed sources. Among them, a substantial sample of high-redshift quiescent galaxies can be identified for follow-up high-resolution spectroscopy.

\section*{Acknowledgement}

SB is supported by the ERC Starting Grant “Red Cardinal”, GA 101076080.

This paper includes data gathered with the 6.5-meter Magellan Telescopes located at Las Campanas Observatory, Chile. Part of the data presented herein were obtained at the W. M. Keck Observatory, which is operated as a scientific partnership among the California Institute of Technology, the University of California and the National Aeronautics and Space Administration. The Observatory was made possible by the generous financial support of the W. M. Keck Foundation. The authors wish to recognize and acknowledge the very significant cultural role and reverence that the summit of Maunakea has always had within the indigenous Hawaiian community.  We are most fortunate to have the opportunity to conduct observations from this mountain. 

\bibliography{Ref}

\bibliographystyle{aasjournal}

\end{document}